\documentclass[aps,prd,onecolumn,groupedaddress,showpacs,nofootinbib,amssymb]{revtex4}
\usepackage{graphicx,bm,color}
\usepackage{amsmath}
\usepackage{amssymb}
\usepackage{amsfonts}

\newcommand{\be}{\begin{equation}}
\newcommand{\ee}{\end{equation}}
\newcommand{\bea}{\begin{eqnarray}}
\newcommand{\eea}{\end{eqnarray}}
\newcommand{\beaa}{\begin{eqnarray*}}
\newcommand{\eeaa}{\end{eqnarray*}}

\newcommand{\tr}{\mathrm{tr}\,}

\allowdisplaybreaks[4]

\begin{document}

\title{Noether current from surface term, Virasoro algebra and black hole entropy in bigravity}
\author{Taishi Katsuragawa$^{1}$ and Shin'ichi Nojiri$^{1,2}$}
\affiliation{
$^1$ Department of Physics, Nagoya University, Nagoya
464-8602, Japan \\
$^2$ Kobayashi-Maskawa Institute for the Origin of Particles and
the Universe, Nagoya University, Nagoya 464-8602, Japan
}



\begin{abstract}
We consider the static, spherically symmetric black hole solutions in the bigravity theory for a minimal model 
with a condition $f_{\mu \nu} = C^{2} g_{\mu \nu}$ and evaluate the entropy for black holes.
In this condition, we show that there exists the Schwarzschild solution for $C^{2} = 1$, 
which is a unique consistent solution.
We examine how the massive spin-$2$ field contributes to and affects the Bekenstein-Hawking entropy 
corresponding to Einstein gravity.
In order to obtain the black hole entropy, 
we use a recently proposed approach, which uses Virasoro algebra 
and central charge corresponding to the surface term in the gravitational action.
\end{abstract}

\pacs{04.60.-m, 04.62.+v}

\maketitle

\section{Introduction}

As is well known, while general relativity and the standard model based on quantum field theory are very successful 
in describing experiments and observations, 
the unification of these two frameworks still remains mysterious because of both conceptual and technical difficulties.
So it is ultimate goal for modern theoretical physics to establish gravitational theory in a microscopic scale, that is, 
the quantum theory of gravity.
One possible way to understand the underlying quantum gravity theory could be to study black hole entropy.

In general relativity, black holes have concepts of temperature and entropy, 
and there exist thermodynamics laws of black holes \cite{Bekenstein:1973ur,Bardeen:1973gs,Hawking:1974rv}.
Investigating the statistical origin of black hole entropy, 
we could be able to obtain quantum properties of space-time, 
and we may have a glimpse of aspects of quantum gravity. 
In fact, the black hole entropy has been calculated in string theory \cite{Strominger:1996sh} 
and loop quantum gravity \cite{Rovelli:1996dv,Ashtekar:1997yu}. 

Recently, a new approach for evaluating the black hole entropy was proposed by using 
the Noether current corresponding to 
the surface term of the action \cite{Majhi:2011ws,Majhi:2012tf,Majhi:2012nq}.
The basic idea of this procedure is the following. 
Define the Noether charges and the Lie bracket corresponding to diffeomorphism; 
then the Lie bracket forms the Virasoro algebra with a central extension, 
from which we can read off a deduced central charge and zero-mode eigenvalues of the Fourier modes of the charge. 
By substituting these quantities into the Cardy formulas \cite{Cardy:1986ie,Bloete:1986qm,Carlip:1998qw}, 
one can obtain the black hole entropy. 
This approach is the general formulation, and it has been shown that the formula of black hole entropy 
can be obtained not only in the Einstein gravity with the usual Einstein-Hilbert action 
but also in higher-curvature gravity \cite{Zhang:2012fq,Kim:2013qra}. 

In this paper, we apply this approach to the bigravity theory.
Bigravity is nonlinear massive gravity that has ghost-free construction 
with the dynamical metric \cite{Fierz:1939ix,Hinterbichler:2011tt,deRham:2010ik,Hassan:2011hr,Hassan:2011zd}. 
This gravity model is called bigravity or bimetric gravity because the model contains two metrics and 
a massive spin-$2$ field appears in addition to the massless spin-$2$ field corresponding to the graviton. 
Therefore, by considering the black holes in bigravity, we can evaluate how the massive spin-$2$ 
field near the horizon affects the black hole entropy \cite{Volkov:2012wp,Banados:2011hk,Banados:2011np}.
For simplicity, we consider the bigravity theory in the minimal model, 
and we obtain entropy of a static, spherically symmetric black hole.

\section{Black hole solution for bigravity}

In this section, we briefly review the construction of ghost-free bigravity \cite{Hassan:2011zd} 
and consider the black hole solutions.
The bimetric gravity includes two metric tensors $g_{\mu \nu}$ and $f_{\mu \nu}$, 
and it contains the massless spin-$2$ field corresponding to the graviton and massive spin-$2$ field. 
It has been shown that the Boulware-Deser ghost does not appear in such a theory.

Note that $g_{\mu \nu}$ and $f_{\mu \nu}$ do not simply correspond to the massless and massive spin-$2$ field individually,
but they are given by linear combinations of these two fields in the linearized model; 
that is, when one considers perturbations of $g_{\mu \nu}$ and $f_{\mu \nu}$ by appropriately redefining two tensor fields, 
the Fierz-Pauli mass term is reproduced.


The action of bigravity is given by
\begin{equation}
S_\mathrm{bigravity} = M^{2}_{g} \int \, d^{4}x \sqrt{-g}R(g) + M^{2}_{f} \int \, d^{4}x \sqrt{-f}R(f) 
+ 2m^{2}_{0} \, M^{2}_\mathrm{eff} \int \, d^{4}x \sqrt{-g} \sum^{4}_{n=0} \beta_{n}e_{n}
\left( \sqrt{g^{-1}f} \right)\, . \label{the action}
\end{equation}
Here $R(g)$ and $R(f)$ are the Ricci scalar for $g_{\mu \nu}$ and $f_{\mu \nu}$, respectively. 
$M_\mathrm{eff}$ is defined by
\begin{equation}
\frac{1}{M^{2}_\mathrm{eff}} = \frac{1}{M^{2}_{g}} + \frac{1}{M^{2}_{f}}\, ,
\label{Meff}
\end{equation}
and $\sqrt{g^{-1}f}$ is defined by the square root of $g^{\mu \rho}f_{\rho \nu}$, that is,
\begin{equation}
\left( \sqrt{g^{-1}f} \right)^{\mu}_{\ \rho} \left( \sqrt{g^{-1}f} \right)^{\rho}_{\ \nu} 
= g^{\mu \rho}f_{\rho \nu}\, .
\label{sqrtfg}
\end{equation}
For general tensor $X^{\mu}_{\ \nu}$, $e_{n}(X)$s are defined by
\begin{eqnarray}
&& e_{0} = 1 \, , \quad 
e_{1} = [X] \, , \quad 
e_{2} = \frac{1}{2} \left( [X]^{2} - [X^{2}] \right) \, , \quad 
e_{3} = \frac{1}{6} \left( [X]^{3} - 3[X][X^{2}] + 2[X^{3}] \right) \, , \nonumber \\
&& e_{4} = \cfrac{1}{24} \left( [X]^{4} - 6[X]^{2}[X^{2}] + 3[X^{2}]^{2} + 8[X][X^{3}] - 6[X^{4}] 
\right) \, , \quad 
e_{k}= 0 \ \ \mbox{for} \ \ k>4 \, .
\end{eqnarray}
Here, $[X]$ expresses the trace of tensor $X$: $[X]=X^{\mu}_{\ \mu}$.

The action for the minimal model is given by
\begin{align}
S_\mathrm{bigravity} =& M^{2}_{g} \int \, d^{4}x \sqrt{-g}R(g) + M^{2}_{f} \int \, d^{4}x \sqrt{-f}R(f) \nonumber \\
& + 2m_{0}^{2} \, M^{2}_\mathrm{eff} \int \, d^{4}x \sqrt{-g} \left( 3 - \tr \sqrt{g^{-1}f} 
+ \det \sqrt{g^{-1}f} \right) \, ,
\label{minimal model}
\end{align}
corresponding to the coefficients,
\begin{equation}
\label{betas}
\beta_{0}=3 \, , \quad \beta_{1}=-1 \, , \quad \beta_{2}=0 \, , \quad \beta_{3}=0 \, , 
\quad \beta_{4}=1\, .
\end{equation}
By the variation (\ref{minimal model}) over $g_{\mu \nu}$ and $f_{\mu \nu}$, we obtain 
\cite{Nojiri:2012zu,Nojiri:2012re}
\begin{align}
0 =& M^{2}_{g} \left( \frac{1}{2}g_{\mu \nu}R(g) - R_{\mu \nu}(g) \right) 
+ m^{2}_{0} \, M^{2}_\mathrm{eff} \left[ \left(3 - \tr \sqrt{g^{-1}f} \right) g_{\mu \nu} \right. \nonumber \\
& \left. 
+ \frac{1}{2}f_{\mu \rho} \left( \sqrt{g^{-1}f} \right)^{-1 \rho}_{\ \ \ \ \nu} 
+ \frac{1}{2}f_{\nu \rho} \left( \sqrt{g^{-1}f} \right)^{-1 \rho}_{\ \ \ \ \mu} \right] \, , \\
0 =& M^{2}_{f} \left( \frac{1}{2}f_{\mu \nu}R(f) - R_{\mu \nu}(f) \right) 
+ m^{2}_{0} \, M^{2}_\mathrm{eff} \sqrt{ \det (f^{-1}g)} \left[ \det (\sqrt{g^{-1}f})f_{\mu \nu} 
\right. \nonumber \\
& \left. 
 - \frac{1}{2} f_{\mu \rho} \left( \sqrt{g^{-1}f} \right)^{\rho}_{\ \nu} 
 - \frac{1}{2} f_{\nu \rho} \left( \sqrt{g^{-1}f} \right)^{\rho}_{\ \mu} \right] \, .
\end{align}
By imposing a condition $ f_{\mu \nu} = C^{2} g_{\mu \nu} $ with a constant $C$, 
the above two equations have the following form:
\begin{align}
0 =& R_{\mu \nu}(g) - \frac{1}{2}g_{\mu \nu}R(g) + \Lambda_{g}g_{\mu \nu} \, ,\label{eom for g} \\ 
0 =& R_{\mu \nu}(f) - \frac{1}{2}f_{\mu \nu}R(f) + \Lambda_{f}f_{\mu \nu} \, .\label{eom for f}
\end{align}
Here, the cosmological constants $\Lambda_{g}$ and $\Lambda_{f}$ are defined by
\begin{align}
\Lambda_{g} \equiv& 3 m^{2}_{0} \left( \frac{M_\mathrm{eff}}{M_{g}} \right)^{2} (|C| - 1) \, ,
\label{lambda_{g}} \\
\Lambda_{f} \equiv& m^{2}_{0} \left( \frac{M_\mathrm{eff}}{M_{f}} \right)^{2} C^{-4}(|C| - C^{4}) \, .
\label{lambda_{f}}
\end{align}
For the consistency, both Eqs. (\ref{eom for g}) and (\ref{eom for f}) should be identical to each other. 
By putting $ f_{\mu \nu} = C^{2} g_{\mu \nu}$, we find $R_{\mu \nu}(f)=R_{\mu \nu}(g)$, $R(f)=C^{-2}R(g)$. 
Then, we find 
\begin{equation}
\Lambda_{g} = C^{2} \Lambda_{f}
\end{equation}
and obtain the constraint for the constant $C$ from Eqs. (\ref{lambda_{g}}) and (\ref{lambda_{f}}): 
\begin{equation}
C^{4} + 3M^{2}_\mathrm{ratio} \, C^{2}|C| - 3M^{2}_\mathrm{ratio} \, C^{2} - |C| = 0 \, .
\label{eqC}
\end{equation}
Here, we define $M_\mathrm{ratio} \equiv M_{f}/M_{g} $.
The solution of Eq. (\ref{eqC}) is given by $C^{2}=1$.
When $C^{2}=1$, the cosmological constants in Eqs. (\ref{lambda_{g}}) and (\ref{lambda_{f}}) vanish, 
and Eqs. (\ref{eom for g}) and (\ref{eom for f}) 
have the identical Schwarzschild solutions,
\begin{eqnarray}
g_{\mu \nu}dx^{\mu}dx^{\nu} = f_{\mu \nu}dx^{\mu}dx^{\nu}
= - (1-\frac{2M}{r}) dt^{2} + \frac{1}{1-\frac{2M}{r}}dr^{2} + r^{2} (d\theta^{2} + \sin^{2} \theta d\phi^{2}) 
\, .
\end{eqnarray}
As a result, we have shown that the bigravity theory for the minimal model has a consistent solution 
when $f_{\mu \nu} = g_{\mu \nu}$ and cosmological constants vanish.
However, for the other model, which has a different choice of $\beta_{n}$s,
it is shown that one can obtain the consistent solutions with a nonvanishing cosmological constant 
\cite{Hassan:2012wr,Hassan:2012rq}. 

Although we can also obtain the Kerr solutions, for simplicity, 
we only consider the Schwarzschild solutions in the following arguments.
Note that when we interpret $g_{\mu \nu}$ as a physical metric, the horizon is at $r=2M$, 
and another tensor field $f_{\mu \nu}$ diverges at the same location, but the divergence can be 
also removed by the coordinate transformation. 

\section{Black hole entropy from the Noether current}

In this section, we summarize the procedure by Majhi 
and Padmanabhan \cite{Majhi:2011ws,Majhi:2012tf,Majhi:2012nq,Zhang:2012fq}.
Let us consider a general surface term as follows: 
\begin{equation}
I_{B} 
= \frac{1}{16 \pi G} \int_{\mathcal{\partial M}} \, d^{n-1}x \sqrt{\sigma} \mathcal{L}_{B} 
= \frac{1}{16 \pi G} \int_{\mathcal{M}} \, d^{n}x \sqrt{g}\nabla_{a}(\mathcal{L}_{B} N^{a})\, .
\end{equation}
Here, $N^{a}$ is a unit normal vector of the boundary $\partial \mathcal{M}$, 
$g_{\mu \nu}$ is the bulk metric, and $\sigma_{\mu \nu}$ is the induced boundary metric.
For the Lagrangian density $ \sqrt{g} \mathcal{L} = \sqrt{g}\nabla_{a}(\mathcal{L}_{B} N^{a})$, 
the conserved Noether current corresponding to differmorphism $ x^{a} \rightarrow x^{a} + \xi^{a} $ 
is given by 
\begin{equation}
J^{a}[\xi] 
= \nabla_{b}J^{ab}[\xi] 
= \frac{1}{16 \pi G} \nabla_{b}
\left[ \mathcal{L}_{B} \left( \xi^{a}N^{b} - \xi^{b}N^{a} \right) \right]\, .
\end{equation}
Here, $J^{ab}$ is the Noether potential, 
and the corresponding charge is defined as
\begin{equation}
Q[\xi] = \frac{1}{2} \int_{\partial \Sigma} \, \sqrt{h} d\Sigma_{ab}J^{ab}\, .
\end{equation}
Here, $ d\Sigma_{ab} = - d^{n-2}x \left( N_{a}M_{b} - N_{b}M_{a} \right)$ is 
the surface element of the $(n-2)$-dimensional 
surface $\partial \Sigma$, and $h_{ab}$ is the corresponding induced metric.
We now choose the unit normal vectors $N^{a}$ and $M^{b}$ as spacelike and timelike, respectively.
In the following disucussion, we assume $\Sigma$ exists near the horizon of a black hole. 

Next, we define the Lie bracket for the charges as follows: 
\begin{equation}
[Q_{1}, Q_{2}] 
= \left( \delta_{\xi_{1}} Q[\xi_{2}] - \delta_{\xi_{2}} Q[\xi_{1}] \right) 
= \int_{\partial \Sigma} \, \sqrt{h} d\Sigma_{ab} 
\left( \xi^{a}_{2} J^{b}[\xi_{1}] - \xi^{a}_{1} J^{b}[\xi_{2}] \right)\, , 
\label{lie bracket}
\end{equation}
which leads to the Virasoro algebra with central extension as we will see later. 
By using the deduced central charge and the Cardy formula, one can find black hole entropy.

To derive the Noether charge and Virasoro algebra, we need to identify appropriate diffeormorphisms, that is, 
the related vector field $\xi^{a}$.
In this work, we consider static-spherical black holes, for which the metric has the following form: 
\begin{equation}
ds^{2} = -f(r)dt^{2} + \frac{1}{f(r)}dr^{2} + r^{2}\Omega_{ij}(x)dx^{i}dx^{j}\, .
\label{BHmetric}
\end{equation}
Here, $\Omega_{ij}(x)$ is the $(n-2)$-dimensional space, 
and $h_{ij}=r^{2}\Omega_{ij}(x)$.
The horizon exist at $r=r_{h}$, where $f(r_{h})=0$.
For the metric (\ref{BHmetric}), the two normal vectors $N^{a}$ and $M^{a}$ are given by 
\begin{equation}
N^{a} = \left( 0, \sqrt{f(\rho + r_{h})}, 0, \cdots , 0 \right) \, , \quad 
M^{a} = \left( \frac{1}{\sqrt{f(\rho + r_{h})}}, 0, \cdots , 0 \right) \, .
\end{equation}
Here, $\rho$ is defined by $r = \rho + h_{h}$ for convenience, and in the near horizon limit, 
we find $\rho \rightarrow 0$.
Then, the metric has the following form:
\begin{equation}
ds^{2} = -f(\rho + r_{h})dt^{2} + \frac{1}{f(\rho + r_{h})}d\rho^{2} 
+ (\rho + r_{h})^{2}\Omega_{ij}(x)dx^{i}dx^{j} \, .
\end{equation}
Furthermore, we introduce the Bondi-like coordinates, 
\begin{equation}
du = dt - \frac{d\rho}{f(\rho + r_{h})}\, , \label{bondi-like coordinate}
\end{equation} 
and the metric is transformed as 
\begin{equation}
ds^{2} = - f(\rho + r_{h})du^{2} - 2dud\rho + (\rho + r_{h})^{2}\Omega_{ij}(x)dx^{i}dx^{j} \, .
\end{equation}
We choose the vector fields $\xi^{a}$ so that the vector fields keep 
the horizon structure invariant.
Then, we now solve the Killing equations for above metric:
\begin{equation}
\mathcal{L}_{\xi} g_{\rho \rho} = -2\partial_{\rho} \xi^{u} = 0 \, ,\quad
\mathcal{L}_{\xi} g_{u \rho}
= - \partial_{u} \xi^{u} - f(\rho + r_{h}) \partial_{\rho} \xi^{u} - \partial_{\rho} \xi^{\rho} = 0\, .
\end{equation}
The solutions of the above equations are given by 
\begin{equation}
\xi^{u} = F(u,x) \, ,\quad 
\xi^{\rho} = - \rho \partial_{u}F(u,x) \, .
\end{equation}
The remaining condition $\mathcal{L}_{\xi}g_{uu}=0$ is automatically satisfied near the horizon because the 
above solutions lead to $\mathcal{L}_{\xi}g_{uu} = \mathcal{O}(\rho)$ and $\rho \rightarrow 0$ at the horizon. 
Expressing the results in the original coordinates $(t, \rho)$, we obtain
\begin{equation}
\xi^{t} = T - \frac{\rho}{f(\rho + r_{h})} \partial_{t}T \, ,\quad 
\xi^{\rho} = - \rho \partial_{t}T \, , \quad 
T(t, \rho, x) = F(u, x) \, .
\end{equation}
Since we have the appropriate vector fields $\xi^{a}$ for a given $T$, we can calculate the charge $Q$.

Finally, expanding $T$ in terms of a set of basis functions $T_{m}$ with
\begin{equation}
T = \sum_{m} A_{m}T_{m} \, ,\quad A^{*}_{m} = A_{-m}\, ,
\end{equation}
we obtain corresponding expansions for $\xi^{a}$ and $Q$ in terms of $\xi^{a}_{m}$ and $Q_{m}$ defined by $T_{m}$.
We choose $T_{m}$ to be the basis so that the resulting $\xi^{a}_{m}$ obeys the algebra isomorphic 
to Diff $S^{1}$, 
\begin{equation}
i\{ \xi_{m}, \xi_{n} \}^{a} = (m-n) \xi^{a}_{m+n}\, ,
\end{equation}
with $\{ , \}$ as the Lie bracket.
Such a $T_{m}$ can be achieved by the choice
\begin{equation}
T_{m} = \frac{1}{\alpha}\mathrm{exp}[im(\alpha t + g(\rho) + p \cdot x)] . \label{T_{m}}\, .
\end{equation}
Here, $\alpha$ is a constant, $p$ is an integer, and $g(\rho)$ is a function that is regular at the horizon.
Note that $\alpha$ is arbitrary in this approach, which will not affect the final results.

\section{Entropy for bigravity}

Using the previous procedure and the black hole solution, we evaluate the black hole entropy.
At first, we need to calculate surface term of the bigravity action $\mathcal{L}_{B}$ 
and the vector field $\xi^{a}$ related to the diffeormophism, which leaves the horizon structure invariant.

Since the interaction term does not include any derivative terms, the contribution to the surface term does not appear.
Therefore, the surface term is two Gibbons-Hawking terms from Ricci scalar $R(g)$ and $R(f)$,
\begin{equation}
\mathcal{L}_{B} = 2K(g) + 2K(f) \, ,
\end{equation}
with $K = - \nabla_{a}N^{a}$ as the trace of the extrinsic curvature of the boundary. 
When we consider the Schwarzschild solution, $f_{\mu \nu} = g_{\mu \nu}$ and $f(r) = 1 - \frac{2M}{r}$ 
with the horizon at $r_{h} = 2M$.
And the metric corresponding to the coordinates $(t, \rho)$ is given by 
\begin{equation}
ds^{2} = - \frac{\rho}{\rho + 2M} dt^{2} + \frac{\rho + 2M}{\rho}d\rho^{2} 
+ (\rho + 2M)^{2}(d\theta^{2} + \sin^{2} \theta d\phi^{2}) \, .
\end{equation}
The Bondi-like coordinate transformation (\ref{bondi-like coordinate}) is defined as
\begin{eqnarray}
du = dt - \frac{2M + \rho}{\rho} \, .
\end{eqnarray}
In this coordinate system, the metric is expressed as 
\begin{equation}
ds^{2} = - \frac{\rho}{\rho + 2M} du^{2} - 2dud\rho 
+ (\rho + 2M)^{2}(d\theta^{2} + \sin^{2} \theta d\phi^{2})\, .
\end{equation}
The vector fields $\xi^{a}$ in the original coordinates $(t, \rho)$ have the following expressions: 
\begin{equation}
\xi^{t} = T - (\rho + 2M) \partial_{t}T \, ,\quad \xi^{\rho} = - \rho \partial_{t}T \, .
\end{equation}

We now calculate the Noether current and the Virasoro algebra.
The normal vectors for the horizon are
\begin{equation}
N^{a} = \left( 0, \sqrt{ \frac{\rho}{\rho + 2M}}, 0, 0 \right) \, , \quad 
M^{a} = \left( \sqrt{\frac{\rho + 2M}{\rho}}, 0, 0, 0 \right) \, ,
\end{equation}
and the Gibbons-Hawking term is given by 
\begin{equation}
K(g) = K(f) = -\frac{2 \rho + M}{\sqrt{\rho} \, (\rho + 2M)^{3/2}}\, .
\end{equation}
The charge $Q$ in the near-horizon limit $\rho \rightarrow 0$ is given by
\begin{equation}
Q[\xi] = 2 \times \frac{1}{8 \pi G} \int_{\mathcal{H}} \, \sqrt{h}d^{2}x 
[\kappa T - \frac{1}{2} \partial_{t}T] \, .
\label{charge}
\end{equation}
Here, $\kappa$ is the surface gravity of the black hole, $\kappa = \frac{1}{4M}$, 
and the factor $2$ comes from the two Gibbons-Hawking terms.

In the calculation, we use the surface gravity $\kappa$, which is related to the expansion 
of $f(\rho + r_{h})$ in the near-horizon limit,
\begin{equation}
f(\rho + r_{h}) = 2\kappa \rho + \frac{1}{2}f''(r_{h})\rho^{2} + \cdots \, , 
\quad \kappa = \frac{f'(r_{h})}{2} .
\end{equation}
Now, we consider $f_{\mu \nu} = g_{\mu \nu}$, so two $\kappa$s from two tensor fields have an identical value.

Finally, we calculate the central charge with the appropriate expansion of $T$.
For $T = T_{m}$, $T_{n}$, the Lie bracket of the charges $Q_{m}$ and $Q_{n}$ (\ref{lie bracket}) 
is given by 
\begin{equation}
[Q_{m}, Q_{n}] = \frac{1}{4 \pi GM}\int_{\mathcal{H}} \, 
\left[ \kappa ( T_{m} \partial_{t}T_{n} - T_{n} \partial_{t}T_{m} ) 
 - \frac{1}{2}( T_{m} \partial^{2}_{t}T_{n} - T_{n} \partial^{2}_{t}T_{m} ) 
+ \frac{1}{4 \kappa} ( \partial_{t} T_{m} \partial^{2}_{t}T_{n} 
 - \partial_{t} T_{n} \partial^{2}_{t}T_{m}) \right] \, .
\label{lie bracket for charge}
\end{equation}
We now substitute $T_{m}$ chosen in the previous section (\ref{T_{m}}) into Eqs.(\ref{charge}) and 
(\ref{lie bracket for charge}) 
and implement the integration over a cross-section area $A$, 
and we obtain 
\begin{eqnarray}
&& Q_{m} = \frac{\kappa A}{4 \pi \alpha G} \delta_{m,0} \, ,\\
&& [Q_{m}, Q_{n}] = - \frac{i \kappa A}{4 \pi \alpha G}(m-n) \delta_{m+n,0} 
 - im^{3} \frac{\alpha A}{8 \pi \kappa G} \delta_{m+n,0} \, .
\end{eqnarray}
Therefore, we find that the central term in the algebra is given by 
\begin{eqnarray}
K[\xi_{m}, \xi_{n}] &=& [Q_{m}, Q_{n}] + i(m-n)Q_{m+n} \nonumber \\
&=& -im^{3} \frac{\alpha A}{8 \pi \kappa G} \delta_{m+n,0} .
\end{eqnarray}
From the central term, we can read off the central charge $C$ 
and the zero mode energy $Q_{0}$ as follows:
\begin{equation}
\frac{C}{12} = \frac{\alpha A}{8 \pi \kappa G} \, , \quad 
Q_{0} = \frac{\kappa A}{4 \pi \alpha G} \, .
\end{equation}
Using the Cardy formula \cite{Cardy:1986ie,Bloete:1986qm,Carlip:1998qw}, 
we eventually obtain the entropy
\begin{equation}
S = 2 \pi \sqrt{\frac{CQ_{0}}{6}} = \frac{A}{2G} \, .
\end{equation}
This is twice as much as the Bekenstein-Hawking entropy in the Einstein gravity.


While we obtain the entropy by using the Noether current corresponding to the surface term, 
it is necessary to check whether our result is appropriate in other approaches.
To compare this result with other methods, we also evaluate the entropy in Wald's approach \cite{Wald:1993nt}.

Here, because $f_{\mu \nu} = g_{\mu \nu}$, the interaction term in Eq.(\ref{minimal model}) vanishes, 
and the Lagrangian depends only on the metric $g_{\mu \nu}$.
Therefore, we may use the Wald formula:
\begin{equation}
S_{\rm Wald}=-2\pi\int_{\mathcal{H}} dA \, \frac{\partial \mathcal{L}}{\partial R_{\alpha \beta \gamma \delta}} 
\epsilon_{\alpha \beta} \epsilon_{\gamma \delta}\, .
\end{equation}
Here, $R_{\alpha \beta \gamma \delta}$ is the Riemann tensor, and $\epsilon_{\alpha \beta}$ is the binormal for the horizon, 
which satisfies the condition $\epsilon_{\alpha \beta} = - \epsilon_{\beta \alpha}$.
For the Schwarzschild solution, we find $\epsilon_{tr} = 1$, and the others are zero.

Then, the Wald entropy is given by
\begin{align}
S =& - 2 \times 2\pi \int_{\mathcal{H}}dA \, \frac{1}{16\pi G} \frac{1}{2}
(g^{\alpha \gamma}g^{\beta \delta} - g^{\beta \gamma}g^{\alpha \delta})\epsilon_{\alpha \beta} \epsilon_{\gamma \delta} \nonumber \\
=& - \frac{1}{2G}\int_{\mathcal{H}}dA \, g^{tt}g^{rr} = \frac{A}{2G}\, ,
\end{align}
which is again twice as much as the Bekenstein-Hawking entropy.

Generally, constructing the Noether current in Wald entropy is rather complicated 
when the $\beta_{n}$s are arbitrary and $f_{\mu \nu}$ expresses the degrees of freedom, which is distinguished with 
$g_{\mu \nu}$.
However, the approach that we use in this paper uses only the surface term to construct the Noether current
in contrast to the Wald entropy, which needs the bulk action.
Therefore, we can ignore the interaction term in evaluating the black hole entropy, 
and the remaining terms corresponding to the Noether current are simply the two traces 
of the extrinsic curvature.
This approach is very simple and useful for calculation.


\section{Conclusion and discussion}

In this work, we have shown that the bigravity for the minimal model has a static, 
spherically symmetric black hole solution.
For the minimal model, we have begun with the ansatz $f_{\mu \nu} = C^{2} g_{\mu \nu}$, but, finally, 
we have shown that the consistency tells $C^2=1$, that is, $f_{\mu \nu}=g_{\mu \nu}$. 
And we have obtained the Schwarzschild solution and evaluated the entropy for it. 

Then, we find that the obtained entropy has a double portion of the Bekenstein-Hawking entropy in the Einstein gravity.
This is because the surface terms for the two tensor fields $f_{\mu \nu}$ and $g_{\mu \nu}$ are identical with each other, 
and they give the same contribution to the Noether current.
Of course, while we have only considered the entropy for the Schwarzschild black hole, 
the stationary axisymmetric solution, that is, the Kerr black hole, is also a solution.
Even in this case, the entropy may have also twice as much as that of the Einstein gravity 
because we have $f_{\mu \nu}=g_{\mu \nu}$ as well.

It is interesting that our approach may be generalized to the case of other models.
For other models, we may choose different values for $\beta_{n}$s in Eq.(\ref{the action}) 
to reproduce the Fierz-Pauli mass term \cite{Damour:2002ws,Hassan:2012wr,Hassan:2012rq}.
We may have other consistent solutions with $f_{\mu \nu}=C^2 g_{\mu \nu}$ but $C^2\neq 1$ in such a model, 
and there could exist (anti-)de Sitter-Schwarzschild solutions and also (anti-)de Sitter-Kerr solutions 
if the cosmological constants do not vanish. 
Forthermore, if two fields are different, the surface terms could give different contributions to 
the Noether current, and the entropy would be changed.

With the procedure using the Noether current of the surface term, we have explicitly shown 
that the entropy is given by sum of two entropies from two Ricci scalars in bigravity.
Similar results are obtain in Refs. \cite{Banados:2011hk,Banados:2011np}, and
our results do not conflict with the implication of general arguments.

\section*{Acknowledgements}

T.K. is partially supported by the Nagoya University Program for Leading Graduate Schools funded by the Ministry 
of Education of the Japanese Government under the program number N01.
The work by S.N. is supported by the JSPS Grant-in-Aid for Scientific Research (S) \# 22224003
and (C) \# 23540296.

\end{document}